\documentclass[twocolumn]{jpsj2} 
\def\runauthor{
\!\!\!\!\!\!\!\!\!\!\!\!\!\!\!\!\!\!\!\!\!\!\!
Yoichi Tanaka, Norio Kawakami, and Akira Oguri
}

\title
{
Numerical Renormalization Group Approach to a Quantum Dot 
Coupled to Normal and Superconducting Leads
}

\author{
Yoichi Tanaka$^1$, Norio Kawakami$^{1,2}$, and Akira Oguri$^3$
}

\inst{
$^{1}$Department of Applied Physics, Osaka University, Suita, Osaka 565-0871, Japan\\
$^{2}$Department of Physics, Kyoto University, Kyoto 606-8502, Japan\\
$^{3}$Department of Material Science, Osaka City University, Osaka 558-8585, Japan
}

\abst{
We study transport through a quantum dot coupled to normal 
and superconducting leads  
using the numerical renormalization group method.
We show that the low-energy properties 
of the system are described by the local Fermi liquid theory 
despite of the superconducting correlations 
penetrated into the dot due to a proximity effect.
We calculate the linear conductance due to the Andreev reflection
in the presence of the Coulomb interaction.
It is demonstrated that the maximum structure 
appearing in the conductance clearly
characterizes a crossover between two distinct spin-singlet
ground states, i.e. the superconducting singlet state and
the Kondo singlet state.
It is further elucidated that the gate-voltage dependence
of the conductance shows different behavior in 
the superconducting singlet region 
 from that in the Kondo singlet region. 
}

\kword{quantum dot, Andreev reflection, Kondo effect,
numerical renormalization group method}

\begin{document}
\maketitle

\section{Introduction}
Recent progress in nano-technology has attracted much interest
in studying quantum transport in mesoscopic systems.
Among others, a quantum dot (QD) \cite{Kouwen}
has played an important role
to reveal correlation effects in the nanoscale systems.
In particular, the observation
\cite{Gold,Cronen}
 of the Kondo effect in QD systems 
\cite{Glaz,Ng,MW,Kawa,Oguri1}
opened a systematic way to investigate strongly correlated electrons, 
which has encouraged 
further theoretical and experimental studies in this field.
Besides this substantial progress, transport properties of 
a mesoscopic system with hybrid normal(metal)-superconductor
junctions have been also investigated extensively.
In this system, the Andreev reflection
\cite{ARreview} plays a key role in physics, 
in which  an incoming
electron from normal side can be reflected as a hole, 
therefore transferring a Cooper pair into superconductor.

The above interesting topics in nanoscale systems 
naturally stimulated the research on the Andreev reflection for a QD
coupled to normal and superconducting leads (N-QD-S)
\cite{Beenakker,Zhao1,Sun1,Loss,Zhao2,Chen}.
In this system, the Andreev reflection (the proximity effect) at the QD-S interface induces the superconducting correlation in the QD, which has a tendency to form the spin-singlet state. On the other hand, for large Coulomb interactions, the Kondo effect is enhanced and therefore the Kondo spin-singlet state is stabilized between the spin moment in the QD and the conduction electrons in the leads. Thus, the competition between these two distinct spin-singlet states occurs in the N-QD-S system. 
In order to clarify how this competition affects the transport in this system, 
theoretical analyses 
\cite{Fazio,Schwab,Clerk,Cuevas,Sun2,Cho,Aono,Avishai1,Avishai2,Krawiec,
Splett,Domanski}
as well as experimental investigations
\cite{Graber}
have been done intensively,
\textit{e.g.} the linear 
\cite{Fazio,Schwab,Clerk,Cuevas,Krawiec,Graber}
or 
nonlinear 
\cite{Fazio,Cho,Avishai1,Aono,Krawiec,Domanski,Graber}
conductance, 
the excess Kondo resonance coming from the novel co-tunneling process
(Andreev-normal co-tunneling)
\cite{Sun2},
the Andreev reflection through a QD embedded in an Aharonov-Bohm ring
\cite{Avishai2} and
the adiabatic electron pumping
\cite{Splett} etc.
In particular, some theoretical studies on the linear conductance have clarified that the Coulomb interaction suppresses the Andreev reflection at the QD-S interface, which leads to the decrease of the linear conductance \cite{Fazio,Schwab,Clerk,Krawiec}. However, these studies have been done on the assumption that the Coulomb interaction in the QD is sufficiently large. 
On the other hand, Cuevas \textit{et al.} analyzed 
the conductance over the range from the noninteracting case 
to the strong-correlation limit,
using a modified second-order perturbation theory \cite{Cuevas}. 
They found that the increase of the coupling
between the QD and the superconducting lead makes it possible
to restore the conductance possibly up to the maximum value $4e^2/h$,
although  physical implications of the conductance maximum
in the presence of the Coulomb interaction
were not discussed in detail.
In an N-QD-S system, moreover, the total number of electrons is not conserved, 
because of the superconducting correlation. 
In such a system, it is not obvious whether the low-energy properties are 
described by the local Fermi liquid theory.


In this paper, we theoretically investigate the transport in an N-QD-S
system with the use of 
the numerical renormalization group (NRG) method
\cite{Wilkins,Hewson1},
 which has been applied successfully to a Josephson current
through a QD
\cite{Choi,Oguri2,Oguri3}.
Applying the Bogoliubov transformation,
we first show  that the low-energy properties 
of an N-QD-S system are described by the local Fermi liquid theory.
Using the NRG method, we calculate the conductance due to 
the Andreev reflection with high accuracy and thus
confirm  Cuevas \textit{et al.}'s results.
To understand the behavior of the conductance,
 the renormalized parameters, which characterize
the Andreev bound states around the Fermi energy, are calculated. From 
the analysis of the ground state properties, 
we demonstrate that the conductance maximum clearly 
characterizes a crossover between 
 the superconducting singlet state and
the Kondo singlet state.

This paper is organized as follows. 
In the next section, we introduce the model and 
describe the low-energy properties
in terms of the local Fermi liquid theory.
Then in \S \ref{sec:result}, we calculate the conductance
and the renormalized parameters.
We discuss how the interplay between the Kondo effect and the superconducting
correlation is reflected in the transport,
by systematically changing the Coulomb interaction, the 
tunneling amplitude and the gate voltage.
A brief summary is given in the
last section.

\section{Model and Formulation} \label{sec:model}

\subsection{Model}
The Hamiltonian of a QD coupled to normal (N) and
superconducting (S) leads is given by
\begin{eqnarray}
H = H_d^0 + H_d^U + H_S + H_N + H_{TS} + H_{TN , } 
\label{Hami1}
\end{eqnarray}
where $H_d^0 + H_d^U$ and $H_{S(N)}$ represent the QD part and
the superconducting (normal) lead part, respectively. 
$H_{TS}$ and $H_{TN}$ are the mixing terms between the QD and the leads.
The explicit form of each part reads 
\begin{eqnarray}
H_d^0 
\!\!\!\!\!&=&\!\!\!\!\! 
\left(\varepsilon_d+\frac{U}{2}\right)(n_d-1),
\,\,\,\,
H_d^U = \frac{U}{2}(n_d-1)^2,
\nonumber\\
H_S 
\!\!\!\!\!&=&\!\!\!\!\!  
\sum_{q,\sigma}\varepsilon _{q}
s_{q\sigma}^\dag s_{q\sigma}^{}
+\sum_{q}(\Delta s_{q\uparrow }^\dag s_{-q\downarrow }^\dag + \textrm{H.c.}),
\nonumber\\
H_N 
\!\!\!\!\!&=&\!\!\!\!\! 
 \sum_{k,\sigma}\varepsilon _{k}
c_{k\sigma}^\dag c_{k\sigma}^{},
\,\,\,\,
H_{TN} =  \sum_{k,\sigma} \frac{V_N}{\sqrt{M_N}}
(c_{k\sigma}^\dag d_{\sigma}^{} + \textrm{H.c.} ),
\nonumber\\
H_{TS} 
\!\!\!\!\!&=&\!\!\!\!\! 
\sum_{q,\sigma} \frac{V_S}{\sqrt{M_S}}
(s_{q\sigma}^\dag d_{\sigma}^{} + \textrm{H.c.} ).
\label{Hamipart}
\end{eqnarray}
The operator $d^{\dag}_{\sigma}$ creates an electron with 
energy $\varepsilon _{d}$ and spin $\sigma$
at the QD, and $n_d=\sum_{\sigma}d^{\dag}_{\sigma}d^{}_{\sigma}$.
Following ref. \citen{Wilkins}, we here write down the QD part $H_d^0 + H_d^U$ in such a way that the energy of the one-electron occupied state ($n_d=1$) is zero. Note that $H_d^0+H_d^U=\varepsilon_d n_d + Un_{d\uparrow}n_{d\downarrow} + const.$, because $(n_d-1)^2=2n_{d\uparrow}n_{d\downarrow}-n_d+1$. In this representation, $H_d^0$ gives the energy shift due to the deviation from the electron-hole symmetric case ($\varepsilon_d+U/2=0 $).
In the Hamiltonian for leads, 
$s_{q\sigma}^\dag(c_{k\sigma}^\dag)$ is the creation operator
of an electron with the energy 
$\varepsilon _{q}(\varepsilon _{k})$
in the superconducting (normal) lead.
In $H_{TS}(H_{TN})$, $V_S(V_N)$ is the tunneling amplitude between the QD
and the superconducting (normal) lead, and
$M_S(M_N)$ is the number of lattice sites 
in the superconducting (normal) lead.
We assume that the superconducting lead is well described 
by the BCS theory with a superconducting gap 
$\Delta=|\Delta|e^{i\phi_S}$,
where $\phi_S$ is the phase of the superconducting gap.

In what follows, we consider the limiting case 
of $|\Delta| \to \infty $. The essential physics of the Andreev reflection,
 which occurs inside the superconducting gap,
is still captured in this limit.
In this situation,  the Hamiltonian \eqref{Hami1} can be reduced 
exactly to an effective
single-channel Hamiltonian (see Appendix \ref{sec:Delta})
\cite{Oguri2,Affleck}
\begin{eqnarray}
H^\mathrm{eff} = H_d^0 + H_d^U + H_d^\mathrm{SC} + H_N + H_{TN}\,,
\label{Hamieff}
\end{eqnarray}
where
\begin{eqnarray}
H_d^\mathrm{SC} = 
\Delta_d\, d^{\dag}_{\uparrow }d^{\dag}_{\downarrow }+\textrm{H.c.}
\label{HamidSC}
\end{eqnarray}
$H_d^\mathrm{SC}$ denotes the effective
onsite superconducting gap at the QD,
\begin{eqnarray}
\Delta_d\equiv \Gamma_S e^{i\phi_S} _{\quad.}
\label{eq:delta_d}
\end{eqnarray}
Notice here that the resonance strength between 
the QD and the superconducting lead is given by
$\Gamma_{S}(\varepsilon)=
\pi \sum_q V_S^{2}\delta(\varepsilon-\varepsilon_{q})/M_{S}$,
which is reduced to an energy-independent
constant $\Gamma_{S}$ in the wide band limit.
For simplicity, we set $\phi_S=0$ in what follows, 
so that $\Delta_d(\equiv \Gamma_S)$ becomes real.
The reduction in the number of the channels gives us a practical 
advantage in the NRG calculations,
because this method works with high accuracy for single-channel systems
while the accuracy becomes rather worse for multi-channel systems.

\subsection{Effective Anderson Hamiltonian with normal leads}
In the NRG method, the normal lead part is transformed into a linear chain
after carrying out a standard procedure of the logarithmic discretization
\cite{Wilkins}.
Then, a sequence of the Hamiltonians is obtained as
\begin{eqnarray}
\mathcal{H}_\mathrm{NRG}^\mathrm{eff}  
\!\!\!\!\!&=&\!\!\!\!\!
\Lambda^{(N-1)/2}
\,  
\left(\, 
      H_d^0 + H_d^U + H_d^\mathrm{SC} 
 + \mathcal{H}_{N} + \mathcal{H}_{TN}
\,\right)_,
\nonumber\\
&&\!\!\!\!\!
\label{eq:NRG_SC_cond}
\end{eqnarray}
where
\begin{eqnarray}
&&
\!\!\!\!\!\!\!\!\!\!
\mathcal{H}_{N} 
\,+\,
\mathcal{H}_{TN}
\nonumber\\
\,\,
&&\,\,
=
\sum_{n=-1}^{N-1}
\sum_{\sigma}
 t_n\, \Lambda^{-n/2} 
\, \left(\,
  f^{\dagger}_{n+1\,\sigma}\,f^{\phantom{\dagger}}_{n \sigma}
 \, + \, 
 f^{\dagger}_{n \sigma}\, f^{\phantom{\dagger}}_{n+1\,\sigma}
 \,\right)_.
\nonumber\\
\label{eq:NRG_H_NT}
\end{eqnarray}
In eq. \eqref{eq:NRG_H_NT}, $f_{-1\sigma} = d_{\sigma}$,
and $f_{n\sigma}$ for $n\geq 0$ is an operator for the conduction 
electron in the normal lead.  The hopping factor  
$t_n$ is defined by $t_{-1} \equiv \widetilde{v} \,\Lambda^{-1/2}$ 
for $n=-1$, where
\begin{eqnarray}
  \widetilde{v}
=\sqrt{ \frac{2\,\Gamma_N D\,A_{\Lambda}}{\pi} },
\,\,
A_{\Lambda}=\frac{1}{2}\, 
\left(\, {1+1/\Lambda \over 1-1/\Lambda }\,\right)
\,\log \Lambda ,
\label{eq:t0}
\end{eqnarray}
and for the conduction band ($n\geq 0$),
\begin{eqnarray}
t_n =  
D\, \frac{1+1/\Lambda}{2}
{ 1-1/\Lambda^{n+1}  
\over  \sqrt{1-1/\Lambda^{2n+1}}  \sqrt{1-1/\Lambda^{2n+3}} 
}_.
\label{eq:tn}
\end{eqnarray}
Here,
$D$ is the half-width of the conduction band, and $\Gamma_N$ 
($=\pi \sum_k V_N^{2}\delta(\varepsilon-\varepsilon_{k})|
_{\varepsilon=\varepsilon_{F}}/M_{N}$)
 is the resonance strength between the QD and the normal lead.
The factor $A_{\Lambda}$ is introduced to compare the discretized model
with the original Hamiltonian \eqref{Hamieff} precisely,
and it behaves as $A_{\Lambda}\to 1$ in the continuum limit
$\Lambda\to 1$
\cite{Wilkins,Sakai}.

In principle, we can carry out the NRG calculation for 
the Hamiltonian \eqref{eq:NRG_SC_cond}.
However, we note that $H_d^\mathrm{SC}$ in eq. \eqref{HamidSC}, 
which represents the effective onsite superconducting gap at the QD, 
mixes states with different particle numbers.
This means that the eigenstates of the 
Hamiltonian \eqref{eq:NRG_SC_cond} can not
be classified in terms of the total number of electrons.
To avoid this inconvenience,
we perform the Bogoliubov transformation 
\cite{Satori}, which is summarized in Appendix \ref{sec:Bogo}.
In the present case 
 the superconducting gap is absent in 
$\mathcal{H}_{N} + \mathcal{H}_{TN}$, so that
the Hamiltonian \eqref{eq:NRG_SC_cond} can be mapped onto
the Anderson model without the onsite superconducting gap,
\begin{eqnarray}
&&
\!\!\!\!\!\!\!\!\!\!\!\!\!\!
\Lambda^{-(N-1)/2}\,
\mathcal{H}_\mathrm{NRG}^\mathrm{eff}   
\nonumber \\
&=&\!\!\!\!
E_d\left(\sum_{\sigma}\gamma_{-1\sigma}^\dag
\gamma_{-1\sigma}^{} -1\right)
+\frac{U}{2}\left(\sum_{\sigma}\gamma_{-1\sigma}^\dag
\gamma_{-1\sigma}^{}-1\right)^2
\nonumber\\
&&\!\!\!
+\sum_{n=-1}^{N-1}\sum_{\sigma}
t_n\, \Lambda^{-n/2}
(\gamma_{n+1\sigma}^\dag \gamma_{n\sigma}^{} + \textrm{H.c.} )
,
\label{Hamitil}
\end{eqnarray}
where
\begin{eqnarray}
E_d=\sqrt{\left(\varepsilon_d+\frac{U}{2}\right)^2+\Delta_d^2}\,\;.
\label{eq:E_d}
\end{eqnarray}
The important point is that the total number of Bogoliubov 
quasiparticles 
$\sum_{n=-1}^{N}\sum_{\sigma}
\gamma_{n\sigma}^\dag \gamma_{n\sigma}^{}$
conserves.
Moreover, 
the Hamiltonian \eqref{Hamitil} is
identical to the ordinary
Anderson model (without superconductivity).
Therefore, the low-energy properties can be described by the local Fermi 
liquid theory,
even though the original model of eq. \eqref{eq:NRG_SC_cond} has
the onsite superconducting gap.
Equation \eqref{Hamitil} has been obtained on the assumption that $|\Delta| \to \infty $. Also for finite $|\Delta|$, the coupling to the normal lead via $\Gamma_N$ could make the low-lying energy states at $|\varepsilon| \lesssim \min(|\Delta|,T_K )$ be described by the Fermi liquid, where $T_K$ is the Kondo temperature in the case of $\Gamma_S=0$. 

There is another point to be mentioned here.
By comparing the first term of the Hamiltonian \eqref{Hamitil}
with $H_d^0$ in eq. \eqref{Hamipart},
we notice that the parameter $\left(\varepsilon_d+U/2\right)$
is replaced by $E_d$.
This means that
the Hamiltonian \eqref{Hamitil} corresponds to 
the Anderson model with the energy level of the impurity site
$\bar{\varepsilon}_d=E_d-U/2$, while the term $(U/2)(n_d-1)^2$
 due to the Coulomb interaction 
remains unchanged.
We discuss the case of  $\Gamma_S\ne0$ ($\Delta_d\ne0$) in this paper,
so that we treat  the reduced Hamiltonian \eqref{Hamitil} with
$E_d>0$ in eq. \eqref{eq:E_d} (so-called asymmetric Anderson model).

\subsection{Local Fermi-liquid description}

We now introduce the Green function to formulate 
the density of states (DOS) of the QD.
Following eq. \eqref{eq:Gd_B} in Appendix \ref{sec:Bogo},
we can write down  the retarded Green function of the QD, after the 
Bogoliubov transformation, as follows,
\begin{eqnarray}
G_{d\uparrow,d\uparrow}^r (\varepsilon )
=
u_d^2  G_{\gamma_{-1}\uparrow,\gamma_{-1}\uparrow}^r (\varepsilon )
+
v_d^2  \bar{G}_{\gamma_{-1}\downarrow,\gamma_{-1}\downarrow}^r
(\varepsilon ).
\label{eq:Green_d}
\end{eqnarray}
Note that the coherent factors $u_d, v_d$ are real
because of $\phi_S=0$.
Using eq. \eqref{eq:Green_d}, the DOS of the QD is given by
\begin{eqnarray}
\rho_{d} (\varepsilon )
\!\!\!\!&=&\!\!\!\!
-\frac{1}{\pi} \textrm{Im}G_{d\uparrow,d\uparrow}^r (\varepsilon )
\nonumber\\
\!\!\!\!&=&\!\!\!\!
-\frac{1}{\pi}
\left\{
u_d^2 \textrm{Im}
  G_{\gamma_{-1}\uparrow,\gamma_{-1}\uparrow}^r (\varepsilon )
+
v_d^2 \textrm{Im}
   \bar{G}_{\gamma_{-1}\downarrow,\gamma_{-1}\downarrow}^r
(\varepsilon )
\right\}_.
\nonumber\\
\label{eq:DOS_d}
\end{eqnarray}
The point we wish to stress is that 
$G_{\gamma_{-1}\uparrow,\gamma_{-1}\uparrow}^r (\varepsilon )$
and
$\bar{G}_{\gamma_{-1}\downarrow,\gamma_{-1}\downarrow}^r
(\varepsilon )$
are derived from the generalized Anderson model
 \eqref{Hamitil}, in which the superconductivity does not show
up explicitly.

 By using the formula \eqref{eq:DOS_d}, we can describe the 
Andreev bound states 
in the QD,
which are induced by the Andreev reflection at the QD-S interface.
We here focus on these states around 
the Fermi energy ($\varepsilon \simeq 0$),  where
the self-energy due to the Coulomb interaction
$\Sigma(\varepsilon)$ is approximately given by 
$\Sigma(\varepsilon) \simeq \Sigma(0)
+
\varepsilon 
\left. 
\partial \Sigma(\varepsilon)/\partial \varepsilon
\right|_{\varepsilon=0 \; }$.
Then, the retarded Green function for electrons 
after the Bogoliubov transformation reads
\begin{eqnarray}
G_{\gamma_{-1}\uparrow,\gamma_{-1}\uparrow}^r (\varepsilon )
\!\!\!\!&=&\!\!\!\!
\frac{1}{\varepsilon-E_d+i\Gamma_N
-\Sigma(\varepsilon)}
\nonumber\\
\!\!\!\!&\simeq&\!\!\!\!
\frac{z}{\varepsilon-\widetilde{E}_d+i\widetilde{\Gamma}_N}_{\;,}
\label{eq:Green_ene0}
\end{eqnarray}
where
\begin{eqnarray}
\widetilde{E}_d=z(E_d+\Sigma(0))
\;,\quad
\widetilde{\Gamma}_N=z\Gamma_N ,
\label{eq:Ed_til}
\\
\qquad
z=
\left(
1-\left. 
\frac{\partial \Sigma(\varepsilon)}{\partial \varepsilon}
\right|_{\varepsilon=0}
\right)^{-1}_{.}
\label{eq:Gamma_til}
\end{eqnarray}
As the retarded Green function for a hole,  
$\bar{G}_{\gamma_{-1}\downarrow,\gamma_{-1}\downarrow}^r
(\varepsilon )$, 
is given by 
$\bar{G}_{\gamma_{-1}\downarrow,\gamma_{-1}\downarrow}^r(\varepsilon )
=
-\left(
G_{\gamma_{-1}\uparrow,\gamma_{-1}\uparrow}^r (-\varepsilon )
\right)^*,$ we end up  with
the  DOS around the Fermi energy,
\begin{eqnarray}
\rho_{d} (\varepsilon )
\simeq
\frac{z}{\pi} 
\left\{
\frac{u_d^2 \widetilde{\Gamma}_N}
{(\varepsilon-\widetilde{E}_d)^2+\widetilde{\Gamma}_N^2}
+
\frac{v_d^2 \widetilde{\Gamma}_N}
{(\varepsilon+\widetilde{E}_d)^2+\widetilde{\Gamma}_N^2}
\right\}_.
\label{eq:DOS_d_ene0}
\end{eqnarray}

We next consider the transport properties in the small bias regime,
which are governed by the Andreev reflection induced inside 
the superconducting gap.
According to  Appendix \ref{sec:deri_Con},
the linear conductance $G_{V=0}=dI/dV|_{V=0}$
at zero temperature is given by
\begin{eqnarray}
G_{V=0}
=
\frac{4e^2}{h}\;
4\frac{\Delta_d^2}{E_d^2}\;
\frac
{ (\widetilde{E}_d/\widetilde{\Gamma}_N)^2 }
{ \{ 1+(\widetilde{E}_d/\widetilde{\Gamma}_N)^2 \}^2  }_.
\label{eq:Condu}
\end{eqnarray}
We see that the conductance is determined by the ratio of 
the renormalized parameters 
$\widetilde{E}_{d}/\widetilde{\Gamma}_{N}$, which
are obtained from the eigenvalues of the Hamiltonian \eqref{Hamitil}
at the fixed point
\cite{Hewson2}.

%
\section{Numerical Results} \label{sec:result}
%

In this section, we discuss transport properties 
for the N-QD-S system at zero temperature.
As mentioned in the previous section,
we assume that the superconducting gap
$|\Delta|$ is sufficiently large ($|\Delta| \to \infty $).
In this case the excited states in the continuum outside the superconducting gap can be neglected. It describes a situation where the superconducting gap is much larger than the characteristic energies of the Andreev reflection.

\subsection{Influence of the Coulomb interaction $U$}

Let us first discuss how the Coulomb interaction $U$
affects the transport properties at zero temperature.
To this end, we explore the detailed properties of 
 the Andreev bound states, which can be 
obtained from the renormalized parameters computed
by means of the NRG method with high accuracy. These renormalized
 parameters also determine the conductance, so that 
we can clarify how  the conductance is controlled by
 the Andreev bound states formed around the Fermi energy.

\begin{figure}[h]
\begin{center}
\includegraphics[scale=0.4]{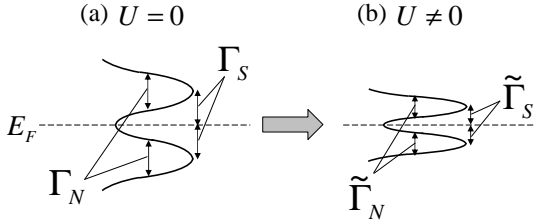}
\end{center}
\caption{Andreev resonances around the Fermi
energy ($E_F=0$) in the symmetric case
($\varepsilon_d+U/2=0$).
}
\label{sketch}
\end{figure}

We observe how the Andreev bound states 
 change their characters 
 with the increase of the Coulomb interaction $U$.
As discussed in \S \ref{sec:model}, the local DOS of the QD 
around the Fermi energy
is given by eq. \eqref{eq:DOS_d_ene0}.
In particular, in the electron-hole symmetric case 
($\varepsilon_d+U/2=0 $),
$E_d=\Gamma_S(\equiv \Delta_d)$ and $u_d^2=v_d^2=1/2$
follow from eqs. 
\eqref{eq:delta_d}, \eqref{eq:E_d} and \eqref{eq:Bogo_factor_B}.
Then, the DOS of the QD, $\rho_{d} (\varepsilon )$, is rewritten as
\begin{eqnarray}
\rho_{d} (\varepsilon )
\simeq
\frac{z}{2\pi} 
\left\{
\frac{\widetilde{\Gamma}_N}
{(\varepsilon-\widetilde{\Gamma}_S)^2+\widetilde{\Gamma}_N^2}
+
\frac{\widetilde{\Gamma}_N}
{(\varepsilon+\widetilde{\Gamma}_S)^2+\widetilde{\Gamma}_N^2}
\right\}_,
\label{eq:DOS_d_ene0_sym}
\end{eqnarray}
where
\begin{eqnarray}
\widetilde{\Gamma}_S=z(\Gamma_S+\Sigma(0)).
\label{eq:GammaS_til}
\end{eqnarray}
 It is seen from eq. \eqref{eq:DOS_d_ene0_sym}
 that $\rho_{d} (\varepsilon )$ in the low-energy region
is determined by the renormalized parameters 
$\widetilde{\Gamma}_S$ and $\widetilde{\Gamma}_N$.
In the noninteracting case ($U=0$), they 
are reduced to the bare ones $\widetilde{\Gamma}_{N}=\Gamma_{N}$ and
$\widetilde{\Gamma}_{S}=\Gamma_{S}$. In this case, 
the Andreev bound states are formed at 
$\varepsilon =\pm\Gamma_S$, which are broadened 
(finite width $\Gamma_N$) by
the coupling between the QD and the normal lead, 
as schematically shown in Fig. \ref{sketch}(a).
We shall refer to these Andreev bound states with finite widths
as the Andreev resonances in the following.
Here we would like to comment on $\Gamma_S$,
which gives the position of the Andreev resonances
shown in Fig. 1(a).
The Andreev reflection at the QD-S interface gives 
rise to the superconducting correlation in the QD.
Taking into account eqs. (4) and (5), in the case of $|\Delta| \to \infty $,
we see the amplitude of the superconducting correlation in the QD is 
given by the resonance strength $\Gamma_S(\equiv \Delta_d)$.
Then, $\Gamma_S$ becomes a parameter indicating the strength of 
the Andreev reflection at the QD-S interface, 
while it originally represents the position of the Andreev resonances 
in the QD.
When the Coulomb interaction $U$ is introduced,
the effective tunneling of electrons between the QD and the leads is 
modified, so that
 the Andreev resonances are renormalized.
Namely, the position and the width of these resonances,
which correspond to the renormalized parameters 
$\widetilde{\Gamma}_S$ and $\widetilde{\Gamma}_N$
respectively, become smaller with the increase of $U$
(see Fig. \ref{sketch}(b)).
Note that when the effect of the Coulomb interaction $U$ is included, 
the strength of the Andreev reflection at the QD-S interface is given by 
$\widetilde{\Gamma}_S$ instead of $\Gamma_S$.
Since we are concerned with the DOS around the Fermi energy, 
the Andreev resonances away from the Fermi energy are not shown
in Fig. \ref{sketch}(b). The overall structure
 including the high energy region
can be found in the literature \cite{Fazio,Clerk,Cuevas,Sun2,Krawiec}.

\begin{figure}[h]
\begin{center}
\includegraphics[scale=0.4]{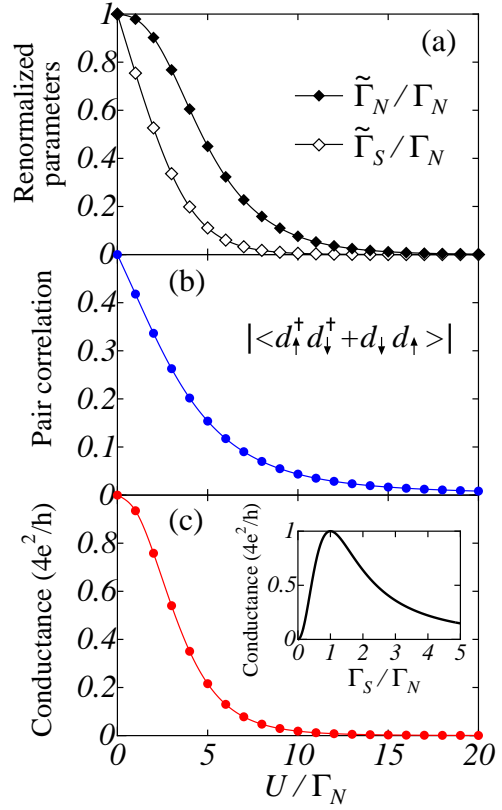}
\end{center}
\caption{
(Color online)
(a)
Renormalized parameters $\widetilde{\Gamma}_S$ and $\widetilde{\Gamma}_N$,
(b)
expectation value of the pair correlation in the QD 
and
(c)
linear conductance as a function of the Coulomb interaction $U$
for $\Gamma_S=\Gamma_N$
in the symmetric case ($\varepsilon_d+U/2=0$).
$\widetilde{\Gamma}_{N(S)}$ and $U$
are normalized by the bare resonance strength, $\Gamma_{N}$.
The NRG calculations have been carried out for 
$\Lambda=3.0$ and $\Gamma_N/D=1.0\times 10^{-3}$.
Inset of (c):
Conductance in the noninteracting case ($U=0$) as a function of
the ratio $\Gamma_S/\Gamma_N$, where we set $\varepsilon_d=0$.
}
\label{reconUr1}
\end{figure}

To observe the formation of the Andreev resonances in our model,
we calculate the renormalized parameters 
$\widetilde{\Gamma}_S$ and $\widetilde{\Gamma}_N$.
Figure \ref{reconUr1}(a) shows the renormalized parameters 
$\widetilde{\Gamma}_{N}$ and $\widetilde{\Gamma}_{S}$ as a function of
the Coulomb interaction $U$ for $\Gamma_S=\Gamma_N$,
where the parameters $\widetilde{\Gamma}_{N(S)}$ 
are normalized by the bare resonance strength, $\Gamma_{N}$.
%
%
As shown in Fig. \ref{reconUr1}(a),
both of $\widetilde{\Gamma}_S$ and $\widetilde{\Gamma}_N$ 
decrease monotonically with the increase of $U$.
The decrease of $\widetilde{\Gamma}_N$ signals
a crossover from the charge-fluctuation regime to the Kondo regime,
as is the case for general N-QD-N systems.
On the other hand, the decrease of $\widetilde{\Gamma}_S$ results 
in the suppression of the Andreev reflection at the QD-S interface,
implying that the superconducting correlation is suppressed in the QD.
To confirm this implication,
we also calculate the expectation value of the pair correlation in the QD,
which is given by
\begin{eqnarray}
\left<
d^{\dag}_{\uparrow }d^{\dag}_{\downarrow }+ 
d^{}_{\downarrow}d^{}_{ \uparrow }
\right>
=-\frac{2}{\pi}
\textrm{tan}^{-1}
(\widetilde{\Gamma}_S/\widetilde{\Gamma}_N).
\label{pairC}
\end{eqnarray}
Note that the sign of 
$\langle
d^{\dag}_{\uparrow }d^{\dag}_{\downarrow }+ 
d^{}_{\downarrow}d^{}_{ \uparrow }
\rangle$ 
depends on the definition of the superconducting gap $\Delta$ in $H_S$:
namely, 
$\langle
d^{\dag}_{\uparrow }d^{\dag}_{\downarrow }+ 
d^{}_{\downarrow}d^{}_{ \uparrow }
\rangle$ 
becomes negative because the second term in the right hand side
of eq. (2) is assumed to be positive.
As shown in Fig. 2(b), 
we see that the absolute value 
$|\langle
d^{\dag}_{\uparrow }d^{\dag}_{\downarrow }+ 
d^{}_{\downarrow}d^{}_{ \uparrow }
\rangle |$ approaches 0 with the increase of $U$ and/or $\widetilde{\Gamma}_S$ in Fig. 2(a).
The Andreev resonances around the Fermi energy are thus renormalized
with the increase of $U$, as shown in Fig. \ref{sketch}.
It is to be noticed that in the large $U$ region where 
the Kondo effect is dominant, these two resonances 
merge around the Fermi energy,  forming a single sharp Kondo resonance,
in accordance with refs. 
\citen{Fazio,Clerk,Cuevas,Sun2} and \citen{Krawiec}.
This is indeed seen from the tendency that  $\widetilde{\Gamma}_S$
decreases more rapidly than $\widetilde{\Gamma}_N$ in Fig. \ref{reconUr1}(a).

We next discuss how the conductance changes
with the increase of $U$. In the symmetric case 
($\varepsilon_d+U/2=0$),
the conductance of eq. \eqref{eq:Condu} is rewritten as
\begin{eqnarray}
G_{V=0}^{\varepsilon_d+\frac{U}{2}=0}
=
\frac{4e^2}{h}\;
\frac
{4 (\widetilde{\Gamma}_S/\widetilde{\Gamma}_N)^2 }
{ \{ 1+(\widetilde{\Gamma}_S/\widetilde{\Gamma}_N)^2 \}^2  }_.
\label{eq:Condu-half}
\end{eqnarray}
The conductance of eq. \eqref{eq:Condu-half} is a function of the ratio $\widetilde{\Gamma}_S/\widetilde{\Gamma}_N$ and has a maximum at $\widetilde{\Gamma}_S/\widetilde{\Gamma}_N=1$, as pointed out in the previous studies \cite{Schwab,Cuevas}.
Here, we would like to mention the relation between the conductance and the Andreev resonances shown in Fig. \ref{sketch}. 
In the case of $\widetilde{\Gamma}_S/\widetilde{\Gamma}_N=1$, which corresponds to the condition for the maximum of the conductance, the distance of the Andreev resonances measured from the Fermi energy ($\widetilde{\Gamma}_S$) is equal to the width of these resonances ($\widetilde{\Gamma}_N$). We also consider the case of $\widetilde{\Gamma}_S/\widetilde{\Gamma}_N\neq1$. When $\widetilde{\Gamma}_S/\widetilde{\Gamma}_N>1$, the distance $\widetilde{\Gamma}_S$ becomes larger than the width $\widetilde{\Gamma}_N$. Then, the DOS of the QD around the Fermi energy gets smaller, which results in the decrease of the conductance. On the other hand, when $\widetilde{\Gamma}_S/\widetilde{\Gamma}_N<1$, the Andreev resonances get closer to the Fermi energy, which leads to the enhancement of the DOS around the Fermi energy. At the same time, however, this means that the transport due to the Andreev reflection is suppressed, so that the conductance decreases as well as the case of $\widetilde{\Gamma}_S/\widetilde{\Gamma}_N>1$. 


Figure \ref{reconUr1}(c) shows the conductance computed as 
a function of the Coulomb interaction $U$ 
for $\Gamma_S=\Gamma_N$.
For reference, we also plot the conductance in the noninteracting case
($U=0$) as a function of the ratio $\Gamma_S/\Gamma_N$
(inset of Fig. \ref{reconUr1}(c)), which is given by replacing 
$\widetilde{\Gamma}_S/\widetilde{\Gamma}_N$ in eq. \eqref{eq:Condu-half} 
with $\Gamma_S/\Gamma_N$.
  From the inset, we see that the conductance for $U=0$ has the maximum 
at $\Gamma_S/\Gamma_N=1$, where
 the couplings to normal and superconducting leads have the same 
amplitude. In this case, the maximum value of the conductance reaches 
the unitary limit $4e^2/h$ because of the electron-hole symmetry.
With the increase of $U$, the conductance decreases monotonically 
 from $4e^2/h$, as shown in Fig. \ref{reconUr1}(c). 
This behavior is in agreement with Cuevas \textit{et al.}'s results,
where they treated a system with 
a finite gap $\Delta$ of the superconducting lead
\cite{Cuevas}.
As stated in ref. \citen{Cuevas},
this is due to the reduction of 
$\widetilde{\Gamma}_S/\widetilde{\Gamma}_N$, which corresponds to the case of 
$\widetilde{\Gamma}_S/\widetilde{\Gamma}_N<1$ stated above.
Similar behavior of monotonic reduction is observed for smaller $\Gamma_S$ 
($\Gamma_S/\Gamma_N<1$).

\begin{figure}[h]
\begin{center}
\includegraphics[scale=0.4]{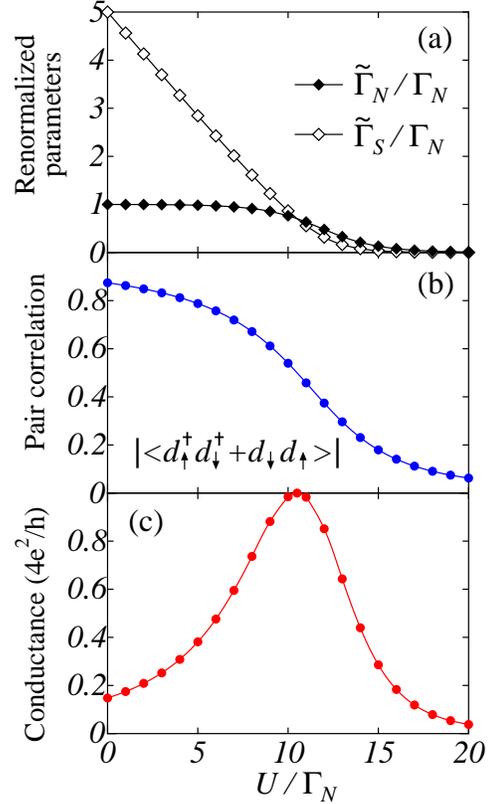}
\end{center}
\caption{
(Color online)
(a)
Renormalized parameters $\widetilde{\Gamma}_S$ and $\widetilde{\Gamma}_N$,
(b)
expectation value of the pair correlation in the QD 
and
(c)
linear conductance as a function of the Coulomb interaction $U$
for $\Gamma_S/\Gamma_N=5$.
The other parameters are the same as in Fig. \ref{reconUr1}.
}
\label{reconUr5}
\end{figure}

Note, however, that somewhat different behavior appears
in the region of $\Gamma_S/\Gamma_N>1$.
Figure \ref{reconUr5} shows similar plots of 
the renormalized parameters and the conductance
for $\Gamma_S/\Gamma_N=5$.
%
%
 Let us focus on the small $U$ region in Fig. \ref{reconUr5}(a) and (b).
In this region, $\widetilde{\Gamma}_S$ decreases monotonically,
which reduces the absolute value
$|\langle
d^{\dag}_{\uparrow }d^{\dag}_{\downarrow }+ 
d^{}_{\downarrow}d^{}_{ \uparrow }
\rangle |$,
as is the case of $\Gamma_S/\Gamma_N=1$.
On the other hand, $\widetilde{\Gamma}_N$ remains almost unchanged
unlike the case of $\Gamma_S/\Gamma_N=1$.
This result indicates that the position of
the Andreev resonances approaches
the Fermi energy, keeping its resonance width unchanged.
However, when the value of $\widetilde{\Gamma}_S$ gets close to
that of $\widetilde{\Gamma}_N$, 
$\widetilde{\Gamma}_N$ also begins to decrease.
For larger $U$, 
both $\widetilde{\Gamma}_S$ and $\widetilde{\Gamma}_N$
approach $0$, as is the case of $\Gamma_S/\Gamma_N=1$.
Comparing $\widetilde{\Gamma}_S$ and $\widetilde{\Gamma}_N$
in Fig. \ref{reconUr5}(a)
with the conductance shown in Fig. \ref{reconUr5}(c),
we see that the conductance has the maximum 
when the condition $\widetilde{\Gamma}_S=\widetilde{\Gamma}_N$
is satisfied.

Summarizing the above results, we can say that
the conductance shows the characteristic $U$-dependence accompanied 
by the maximum structure,
around which the strength of  two effective resonances is exchanged:
we have  $\widetilde{\Gamma}_S > \widetilde{\Gamma}_N$ 
in the small $U$ region ($U/\Gamma_N<10$),
while $\widetilde{\Gamma}_S < \widetilde{\Gamma}_N$
 in the large $U$ region ($U/\Gamma_N>10$).
In the following, we demonstrate that  
the maximum of the conductance characterizes
a crossover between the two distinct singlet regions
where either the superconducting correlation or the Kondo correlation
is dominant.


\subsection{Implications of the conductance maximum}
To discuss the condition that  the conductance
takes the maximum value in more detail,
we would like to observe how the conductance changes
as a function of the resonance strength $\Gamma_S$ 
that gives a measure of the superconducting correlation in the QD,
as has been done in ref. \citen{Cuevas}. We find it more instructive to
 summarize our results  in two-dimensional plots 
 in the plane of $\Gamma_S$ and $U$, which indeed  allows
 us to elucidate how the electron correlations affect 
the conductance maximum.

\begin{figure}[h]
\begin{center}
\includegraphics[scale=0.68]{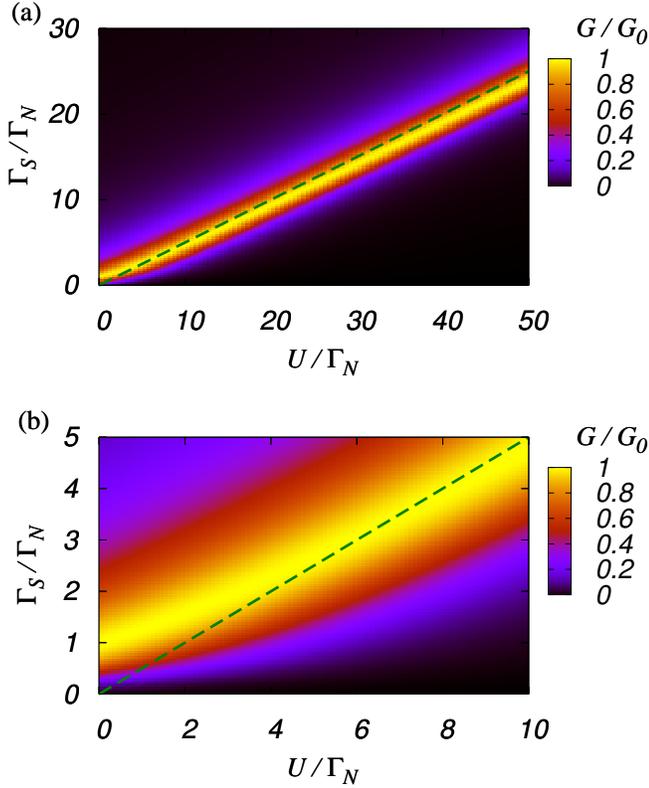}
\end{center}
\caption{
(Color online)
(a)
Color-scale representation of the conductance $G/G_0$,
where $G_0=4e^2/h$, 
as a function of $U/\Gamma_N$ and $\Gamma_S/\Gamma_N$.
The dashed line, $\Gamma_S=U/2$, represents the 
boundary of the singlet-doublet transition 
for $\Gamma_N=0$.
(b)
Enlarged picture in the  region with
small $U/\Gamma_N$ and $\Gamma_S/\Gamma_N$.
}
\label{maxplot}
\end{figure}

Figure \ref{maxplot} (a) is the color-scale representation of the conductance
as a function of $U/\Gamma_N$ and $\Gamma_S/\Gamma_N$.
We also show  the enlarged picture 
in small $U$ and $\Gamma_S$ region in Fig. \ref{maxplot}(b).
As shown in Fig. \ref{maxplot}(b),
the conductance takes the maximum
at $\Gamma_S/\Gamma_N=1$ when $U/\Gamma_N=0$, 
in accordance with the inset of Fig. \ref{reconUr1}.
As $U/\Gamma_N$ increases, 
the value of $\Gamma_S/\Gamma_N$ giving
the maximum of the conductance
increases linearly along the line of $\Gamma_S=U/2$
(see Fig. \ref{maxplot}).
As mentioned above, the maximum of the conductance reaches 
the unitary limit ($G_0=4e^2/h$) because of the electron-hole symmetry.
For instance, around $U/\Gamma_N=10$ and $\Gamma_S/\Gamma_N=5$,
the conductance reaches  the unitary limit,
as already shown in  Fig. \ref{reconUr5}(c).

Here we consider the physical implication of $\Gamma_S=U/2$,
which coincides with the condition that the conductance reaches 
the unitary limit for large values of $\Gamma_S/\Gamma_N$ and $U/\Gamma_N$.
To this end, let  us examine the limit of $\Gamma_N\to 0$.
When $\Gamma_N=0$, the QD is disconnected from the normal lead,
and only connected to the superconducting lead (QD-S).
Recall here that the QD-S system is equivalent to
a magnetic impurity model embedded in a superconductor,
which has been studied for dilute magnetic alloys
\cite{Soda,Shiba,Muller,Matsuura}.
As discussed in refs. \citen{Soda,Shiba,Muller,Matsuura},
the ground state of the QD-S system is a nonmagnetic singlet 
or a magnetic doublet.
Specifically in the limit of $|\Delta| \to \infty $,
the ground state only depends on  the ratio of $\Gamma_S/U$.
Namely, the ground state is 
a superconducting spin-singlet state for $\Gamma_S/U>0.5$
or a spin-doublet state for $\Gamma_S/U<0.5$.
A  transition between these ground states occurs
at $\Gamma_S/U=0.5$.

We now observe what happens for finite $\Gamma_N$.
When the coupling between the QD and the normal lead is introduced,
conduction electrons in the normal lead 
screen the free spin moment to form the Kondo singlet state
in the region of $ \Gamma_S/U<0.5$.
Therefore, the ground state of the N-QD-S system
is always singlet.  There are, however, two distinct singlets,
i.e. one with  superconducting-singlet character for $\Gamma_S/U>0.5$ 
and  the other with Kondo-singlet character for $\Gamma_S/U<0.5$.
 From the above discussion,
we see that the maximum of the conductance in Fig. \ref{maxplot}
for large values of $\Gamma_S/\Gamma_N$ and $U/\Gamma_N$ clearly
characterizes the crossover between these two different spin-singlet states.
We have drawn this conclusion on the assumption that
the gap is sufficiently large ($|\Delta| \to \infty $).
 We believe that this conclusion also holds for
 an N-QD-S system
with a finite gap $\Delta$ of the superconducting lead,
though the onset of the transition of the ground state for $\Gamma_N=0$
depends also on $\Delta$
\cite{Oguri2,Oguri3}.

\begin{figure}[h]
\begin{center}
\includegraphics[scale=0.38]{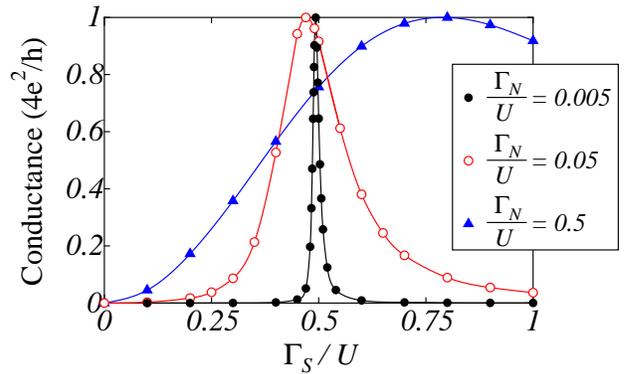}
\end{center}
\caption{
(Color online)
Conductance as a function of
the ratio $\Gamma_S/U$ for
several values of $\Gamma_N/U$,
where the Coulomb interaction
is fixed as $U/D=2.0\times10^{-2}$.
}
\label{congs}
\end{figure}

Before closing this subsection,  we display the 
conductance in Fig. \ref{congs} in a slightly different way
to make the comparison with experiments easier;
it is plotted as a function of the ratio
$\Gamma_S/U$ for several values of $\Gamma_N/U$, 
which are both controllable by 
changing the voltage between the QD and the leads experimentally
\cite{Gold,Cronen}.
 From Fig. \ref{congs},
we see that as $\Gamma_N/U$ gets smaller, 
the peak structure  becomes sharper and  
its position approaches $\Gamma_S/U=0.5$.
Thus, the peak structure in the conductance for $\Gamma_N/U=0.05$ and $0.005$
in Fig. \ref{congs} clearly characterizes the crossover between 
the superconducting and Kondo spin-singlet states, as discussed above.
On the other hand, as $\Gamma_N/U$ becomes larger,
the peak structure is somewhat broadened and its position shifts 
toward $\Gamma_S/U>0.5$,
as shown for $\Gamma_N/U=0.5$
in Fig. \ref{congs}.
Note here that the increase of $\Gamma_N/U$
enhances charge fluctuations in the QD,
making  the Kondo correlation weak. 
Therefore, the crossover between the superconducting and Kondo singlet states
becomes smeared, and accordingly 
the position of  the conductance maximum
deviates from $\Gamma_S/U=0.5$.

\subsection{Away from the symmetric case}

Finally, we discuss the conductance as a function of
the energy level in the QD, $\varepsilon_d$,
away from the electron-hole symmetric case
($\varepsilon_d+U/2\neq 0$).
Note that the energy level in the QD can be changed 
by the gate-voltage control.
A deviation from the electron-hole symmetric case is represented by
$H_d^0$ in eq. \eqref{Hamipart}, as stated in \S \ref{sec:model}.
In this case,
the position of the Andreev resonances is given by
$\widetilde{E}_d$ in eq. \eqref{eq:Ed_til} instead of $\widetilde{\Gamma}_S$.
Following the way outlined in the symmetric case, 
we discuss the characteristics of the conductance
in connection with the ground-state nature of the system.

\begin{figure}[h] 
\begin{center}
\includegraphics[scale=0.75]{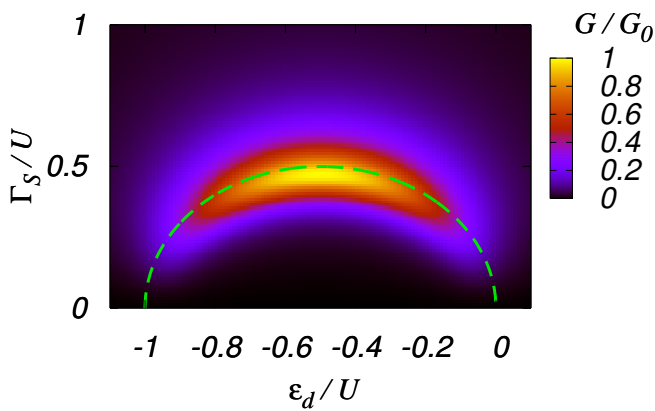}
\end{center}
\caption{
(Color online)
Color-scale representation of the conductance $G/G_0$ ($G_0=4e^2/h$) 
as a function of $\varepsilon_d/U$ and $\Gamma_S/U$.
In this calculation, we set 
$\Gamma_N/U=0.05$ and $U/D=2.0\times10^{-2}$.
The dashed line is 
$\{\left(\varepsilon_d/U+1/2\right)^2+
(\Gamma_S/U)^2\}^{1/2}=1/2$, which gives 
the boundary of the singlet-doublet 
transition for $\Gamma_N=0$.
}
\label{maxploted}
\end{figure}

In Fig. \ref{maxploted} we show the conductance 
in the color-scale representation
as a function of $\varepsilon_d/U$ and $\Gamma_S/U$.
 We also draw a half circle given by
$\{\left(\varepsilon_d/U+1/2\right)^2+(\Gamma_S/U)^2\}^{1/2}=1/2$,
which denotes the boundary of the singlet-doublet transition 
 for $\Gamma_N=0$.
The inside of the half circle is the spin-doublet region at $\Gamma_N=0$,
which is replaced with the Kondo spin-singlet region
in the presence of any finite $\Gamma_N$.
On the other hand, 
the outside is the superconducting spin-singlet region. 
  From Fig. \ref{maxploted}, 
we see that the conductance for $\Gamma_S/U<0.5$ has large 
values along the half circle.
This means that the conductance as a function of $\varepsilon_d$
has two peaks \cite{Cuevas}.
Note here that a deviation from the 
electron-hole symmetric case ($\varepsilon_d/U \neq -0.5$)
makes the Kondo correlation weaker.
Therefore, the two-peak structure of the conductance for $\Gamma_S/U<0.5$
indicates the crossover of  superconducting-Kondo-superconducting 
singlet states.  On the other hand, for $\Gamma_S/U>0.5$,
the ground state is the superconducting singlet state
for any values of $\varepsilon_d/U$.
As seen in eqs. \eqref{eq:E_d} and \eqref{eq:Ed_til}, 
a deviation from the symmetric case drives 
the position of the Andreev resonances ($\widetilde{E}_d$)
away from the Fermi energy.
In the superconducting singlet region ($\Gamma_S/U>0.5$), 
$\widetilde{E}_d/\widetilde{\Gamma}_{N}>1$
is satisfied, so that the deviation leads to the reduction of 
the conductance,
as is the case of $\widetilde{\Gamma}_S/\widetilde{\Gamma}_N>1$ 
in the symmetric case.
Thus, the conductance as a function of $\varepsilon_d$
has a single maximum at $\varepsilon_d/U=-0.5$,
although this peak is not a sign of the crossover between two singlet states
unlike the peaks for $\Gamma_S/U<0.5$.

\section{Summary}

We have investigated transport properties of an N-QD-S system
using the NRG method.
Especially, we have focused on the limiting case of $|\Delta| \to \infty $ 
with particular emphasis on the Andreev
reflection which arises inside the superconducting gap.
Adapting the Bogoliubov transformation to the
simplified model, we have first demonstrated
that our system with superconductivity can be mapped on an effective 
Anderson impurity model without superconductivity, which enables us to
describe the low-energy properties in terms
of the local Fermi liquid theory.

To clarify the influence of the Coulomb interaction
on the transport due to the Andreev reflection,
we have calculated the conductance as a function of
the Coulomb interaction $U$.
For the ratio $\Gamma_S/\Gamma_N>1$, the conductance
has the maximum in its $U$-dependence
while for $\Gamma_S/\Gamma_N \le 1$ it decreases monotonically,
which is in accordance with Cuevas \textit{et al.}'s results.
We have also calculated the renormalized parameters
to discuss the conductance in connection with
 the Andreev resonances around the Fermi energy.
Through the analysis using the renormalized parameters,
 we have found that the maximum of 
the conductance gives an indication of the crossover between
two distinct types of singlet ground states.
To observe the nature of the crossover in detail, we have studied 
the transport properties  by focusing on
the changes of the Coulomb interaction $U$
and the resonance strength $\Gamma_S$.
In particular, starting from the special case with $\Gamma_N=0$,
i.e. the QD system only coupled to the superconducting lead,
we have shown that the conductance 
maximum clearly characterizes the crossover between 
 the Kondo singlet state and the superconducting singlet 
state.

It has been further elucidated that 
the gate-voltage dependence of the conductance
shows different behavior depending on the value of $\Gamma_S$;
 there are two peaks in the gate-voltage dependence
characterizing the crossover of 
superconducting-Kondo-superconducting singlet regions 
for $\Gamma_S/U < 0.5$, whereas
only a single maximum appears in the superconducting
singlet region for $\Gamma_S/U >0.5$.

 In this paper, we have clarified several characteristic transport properties on the assumption that the superconducting gap is sufficiently large. Nevertheless we believe that our main conclusion, {\it i.e.} the conductance maximum clearly characterizes the crossover between two distinct types of singlet ground states, holds even for general N-QD-S systems with a finite gap $\Delta$. In this case, however, the crossover between two distinct types of singlet states depends also on $\Delta$, as stated in \S \ref{sec:result}. In near future, we expect that the characteristic maximum structure found in the conductance due to the Andreev reflection will be observed experimentally.


\section*{Acknowledgment}

We would like to thank Y. Nisikawa and T. Suzuki for valuable discussions.
A part of computations was done at the
Supercomputer Center at the Institute for Solid State
Physics, University of Tokyo.
The work was partly supported by a Grant-in-Aid from the
 Ministry of Education, Culture,
Sports, Science and Technology of Japan.
Y. Tanaka is supported by JSPS Research Fellowships for
Young Scientists.
A. Oguri is supported by a Grant-in-Aid from JSPS.

\appendix

\section{Hamiltonian at $|\Delta| \to \infty $} \label{sec:Delta}

We explain how the Hamiltonian \eqref{Hami1} can be reduced to 
an effective single-channel Hamiltonian \eqref{Hamieff}
in the limit of $|\Delta| \to \infty $.
For this purpose, we consider the Green function of the QD
in the $2\times 2$ Nambu representation,
which takes the form 
\begin{eqnarray}
\textbf{G}_{d,d}(t-t')
\!\!\!\!\!&=&\!\!\!\!\! 
-i\left(
    \begin{array}{cc}
      \langle Td_{\uparrow} (t) d_{\uparrow}^\dag (t') \rangle &
      \langle Td_{\uparrow} (t) d_{\downarrow} (t') \rangle \\
      \langle Td_{\downarrow}^\dag (t) d_{\uparrow}^\dag (t') \rangle &
      \langle Td_{\downarrow}^\dag (t) d_{\downarrow} (t') \rangle
    \end{array}
  \right)
\nonumber\\
\!\!\!\!\!&=&\!\!\!\!\! 
\left(
   \begin{array}{cc}
    G_{d\uparrow,d\uparrow} (t-t') & \bar{F}_{d\uparrow,d\downarrow} (t-t') \\
    F_{d\downarrow,d\uparrow} (t-t') & \bar{G}_{d\downarrow,d\downarrow} (t-t')
   \end{array}
 \right)_.
\nonumber\\
\label{eq:Green_A}
\end{eqnarray}
We perform the Fourier transformation of
the retarded (advanced) Green function for 
eq. \eqref{eq:Green_A}.
 From the Dyson equation, we have
\begin{eqnarray}
\left(
  \varepsilon \; \textbf{I}
  -\varepsilon_d \mbox{\boldmath $\tau_3$}
 -\mbox{\boldmath $\Sigma$}^{r(a)} (\varepsilon)
\right)
\textbf{G}_{d,d}^{r(a)} (\varepsilon)
=
\textbf{I}
\, .
\label{eq:Dyson_A}
\end{eqnarray}
In the interacting case, the self-energy 
$\mbox{\boldmath $\Sigma$}^{r(a)} (\varepsilon)$ 
can be classified into two parts
\begin{eqnarray}
\mbox{\boldmath $\Sigma$}^{r(a)} (\varepsilon)
=
\mbox{\boldmath $\Sigma$}^{r(a)}_0 (\varepsilon)
+
\mbox{\boldmath $\Sigma$}^{r(a)}_U (\varepsilon)
\, ,
\label{eq:Sigma_A}
\end{eqnarray}
where $\mbox{\boldmath $\Sigma$}^{r(a)}_0 (\varepsilon)$ 
is the self-energy due to 
the mixing between the QD and the leads, 
and $\mbox{\boldmath $\Sigma$}^{r(a)}_U (\varepsilon)$ 
is due to the Coulomb interaction.
The exact expression for
$\mbox{\boldmath $\Sigma$}^{r}_0$ ($=(\mbox{\boldmath $\Sigma$}^{a}_0)^\dag$)
is written down as
\begin{eqnarray}
\mbox{\boldmath $\Sigma$}^{r}_0 (\varepsilon)
=
\left(
\begin{array}{cc}
 -i(\Gamma_{N}+\Gamma_{S}\beta(\varepsilon))
&
 i\Gamma_{S}\beta(\varepsilon)\frac{\Delta}{\varepsilon}
\\
 i\Gamma_{S}\beta(\varepsilon)\frac{\Delta^*}
 {\varepsilon}
&
 -i(\Gamma_{N}+\Gamma_{S}\beta(\varepsilon))
\end{array}
\right)_,
\!\!\!\!\!\! \nonumber\\
\label{eq:Sigma0r_A}
\end{eqnarray}
where
\begin{eqnarray}
\beta(\varepsilon)
=
\frac{|\varepsilon|\, \theta(|\varepsilon|-|\Delta|)}
{\sqrt{\varepsilon^2-|\Delta|^2}}
+
\frac{\varepsilon\, \theta(|\Delta|-|\varepsilon|)}
{i\sqrt{|\Delta|^2-\varepsilon^2}}_.
\end{eqnarray}
Here, we consider the case that the superconducting gap 
is sufficiently large, i.e. the limit of $|\Delta| \to \infty $.
In this limit, taking into account
$\beta(\varepsilon)
\to
\frac{\varepsilon}{i\sqrt{|\Delta|^2-\varepsilon^2}}$, 
$\mbox{\boldmath $\Sigma$}^{r}_0 (\varepsilon)$ 
is reduced to
\begin{eqnarray}
\lim_{|\Delta|\to \infty }
\mbox{\boldmath $\Sigma$}^{r}_0 (\varepsilon)= 
\left(
\begin{array}{cc}
-i\Gamma_{N}  &   \Gamma_{S}e^{i\phi_S}        \\
\Gamma_{S}e^{-i\phi_S}         &  -i\Gamma_{N}
\end{array}
\right)_.
\label{eq:Sigma0_red_A}
\end{eqnarray}
We note that the off-diagonal element in eq. \eqref{eq:Sigma0_red_A}
 can be regarded as a static superconducting gap
induced at the QD, and its amplitude is given by 
the resonance strength $\Gamma_S$.
Therefore, in the limit of $|\Delta| \to \infty $ the information about
the superconducting lead can be included through the off-diagonal term
of the Green function of the QD.
This enables us to rewrite the Hamiltonian \eqref{Hami1} to an effective 
single-channel Hamiltonian with an extra superconducting gap at the QD
in  eqs. \eqref{Hamieff} and \eqref{HamidSC}.


\section{Bogoliubov transformation} \label{sec:Bogo}

We perform the Bogoliubov transformation for the Hamiltonian 
\eqref{eq:NRG_SC_cond}.
Using the Nambu representation, $H_d^0 + H_d^\mathrm{SC}$ and 
$\mathcal{H}_{N}+\mathcal{H}_{TN}$ in the Hamiltonian 
\eqref{eq:NRG_SC_cond}
are rewritten as,
\begin{eqnarray}
H_d^0 + H_d^\mathrm{SC}
\!\!\!\!\!&=&\!\!\!\!\!  
 \mbox{\boldmath $\psi$}_{f,-1}^{\dag}
\left(
 \begin{array}{cc}
  \xi_d & \Delta_d^{\phantom{*}} \\
  \Delta_d^* &   -\xi_d 
 \end{array}
\right)
 \mbox{\boldmath $\psi$}_{f,-1 \,\, ,}^{}
\label{eq:Hamid0_B}
\\
\mathcal{H}_{N}
+\mathcal{H}_{TN} 
\!\!\!\!\!&=&\!\!\!\!\! 
\sum_{n=-1}^{N-1}
\!\!
t_n \Lambda^{-n/2}
\left(
\mbox{\boldmath $\psi$}_{f,n+1}^{\dag} \mbox{\boldmath $\psi$}_{f,n}^{}
+\mbox{\boldmath $\psi$}_{f,n}^{\dag} \mbox{\boldmath $\psi$}_{f,n+1}^{}
\right)_,
\nonumber\\
\label{eq:HamiNT_B}
\end{eqnarray}
where
\begin{eqnarray}
\mbox{\boldmath $\psi$}_{f,n}^{}
=
\left(
\begin{array}{c}
  f_{n\uparrow}^{\phantom{\dagger}} \\
 (-1)^{n-1} f_{n\downarrow}^{\dagger} 
\end{array} 
 \right)_.
\label{eq:Hamid0_Bpsi}
\end{eqnarray}
In eq. \eqref{eq:Hamid0_B}, 
$\xi_d = \epsilon_d \,+ \,U/2$ and $f_{-1\sigma} = d_{\sigma}$.
For eqs. \eqref{eq:Hamid0_B} and \eqref{eq:HamiNT_B},
we carry out the Bogoliubov transformation,
\begin{eqnarray}
\mbox{\boldmath $\psi$}_{\gamma,n}^{}
=
\mbox{\boldmath $U$}_{d}^{\dagger} 
\,
\mbox{\boldmath $\psi$}_{f,n \,\, ,}^{}
\label{eq:Bogo_B}
\end{eqnarray}
where
\begin{eqnarray}
&
\!\!\!\!\!\!\!\!\!\!
\mbox{\boldmath $\psi$}_{\gamma,n}^{}
=
\left(
\!\!
 \begin{array}{c}
  \gamma_{n\uparrow}^{\phantom{\dagger}} \\
  (-1)^{n-1} \gamma_{n\downarrow}^{\dagger} 
 \end{array}
\!\!
\right), \,\,
  \mbox{\boldmath $U$}_{d} =
     \left( 
     \!
        \begin{array}{cc}
          u_{d}  &   -v_{d}^* \\
          v_{d}  &   \phantom{-}u_{d}^* 
        \end{array} 
     \!
     \right), 
\label{eq:Bogo_BU} \\
  & |u_d|^2=\frac{1}{2}\left(1+\frac{\xi_d}{E_d}\right), \quad
    |v_d|^2=\frac{1}{2}\left(1-\frac{\xi_d}{E_d}\right), 
\label{eq:Bogo_factor_B} \\
& E_d=\sqrt{\xi_d^2+|\Delta_d^{\phantom*}|^2 }_.
\end{eqnarray}
As a result, each term of
the Hamiltonian $\mathcal{H}_\mathrm{NRG}^\mathrm{eff}$ 
can be transformed as follows,
\begin{eqnarray}
H_d^0 + H_d^\mathrm{SC} 
\!\!\!\!\!&=&\!\!\!\!\! 
E_d \, 
\left( 
\sum_{\sigma}
 \gamma_{-1\sigma}^{\dagger}\, 
\gamma_{-1\sigma}^{\phantom{\dagger}}
 -1 \right)  \;,
\label{eq:Hamid0til_B}
\\
H_{d}^U 
\!\!\!\!\!&=&\!\!\!\!\! 
\frac{U}{2} \, 
\left( 
\sum_{\sigma}
 \gamma_{-1\sigma}^{\dagger}\, 
\gamma_{-1\sigma}^{\phantom{\dagger}}
 -1 \right)^2  \;,
\label{eq:HamidUtil_B}
\\
\mathcal{H}_{N}
+\mathcal{H}_{TN}
\!\!\!\!\!&=&\!\!\!\!\! 
\sum_{n=-1}^{N-1} \,
\sum_{\sigma}
 t_n\, \Lambda^{-n/2} 
\, \left(\,
  \gamma^{\dagger}_{n+1\,\sigma}\,\gamma^{\phantom{\dagger}}_{n \sigma}
 \, + \, 
 \textrm{H.c.}
 \,\right)_.
\label{eq:HamiNTtil_B}
\nonumber\\
\end{eqnarray}
eqs. \eqref{eq:Hamid0til_B}, \eqref{eq:HamidUtil_B} and \eqref{eq:HamiNTtil_B}
 give the Hamiltonian \eqref{Hamitil}.
We should notice that $H_{d}^U$ remains unchanged 
under the Bogoliubov transformation.

We finally show how the Green function of the QD is transformed
via the Bogoliubov transformation.
Applying the Bogoliubov transformation of eq. \eqref{eq:Bogo_B}
to the Fourier transformation of eq. \eqref{eq:Green_A},
\begin{eqnarray}
\textbf{G}_{d,d}(\varepsilon)
=\mbox{\boldmath $U$}_d \>
\textbf{G}_{\gamma_{-1} ,\gamma_{-1} }(\varepsilon) \>
\mbox{\boldmath $U$}_{d\, ,}^\dag
\label{eq:Gd_B}
\end{eqnarray}
where the Green function 
$\textbf{G}_{\gamma_{-1}  ,\gamma_{-1} }(\varepsilon)$
is for the Hamiltonian after the Bogoliubov transformation.
Therefore,  we end up with
$\textbf{G}_{\gamma_{-1} ,\gamma_{-1} }(\varepsilon)$ without
 off-diagonal elements,
\begin{eqnarray}
\textbf{G}_{\gamma_{-1} ,\gamma_{-1} }(\varepsilon)
=
\left(
  \begin{array}{cc}
    G_{\gamma_{-1} \uparrow ,\gamma_{-1} \uparrow }(\varepsilon) & 0 \\
    0 & \bar{G}_{\gamma_{-1} \downarrow ,\gamma_{-1} \downarrow }(\varepsilon)
  \end{array}
\right)_.
\nonumber\\
\label{eq:Ggamma_B}
\end{eqnarray}

\section{Derivation of eq. \eqref{eq:Condu} } \label{sec:deri_Con}

By applying the Landauer formula,
we derive the linear conductance $dI/dV|_{V=0}$.
The current from the normal lead to the QD is given 
in terms of the Nambu representation
\cite{Sun1,Fazio}
\begin{eqnarray}
I
\!\!\!\!\!&=&\!\!\!\!\! 
\frac{2e}{h}
  \int d\varepsilon (i\Gamma_N)
   \left[ \textbf{G}_{d,d}^{r} (\varepsilon)
           \mbox{\boldmath $\Omega$} (\varepsilon)
           \textbf{G}_{d,d}^{a} (\varepsilon)
   \right.
\nonumber\\ 
&& \left. \qquad\quad
  -2i(1-2f_N)
  \textrm{Im} \textbf{G}_{d,d}^{r} (\varepsilon)
   \right]_{11 \; .}      
\label{eq:current_C}
\end{eqnarray}
The subscript $11$ means the (1,1) element of the matrix.
$\mbox{\boldmath $\Omega$}(\varepsilon)$ is the self-energy 
 given by 
$\mbox{\boldmath $\Omega$}(\varepsilon) = 
- \mbox{\boldmath $\Sigma$}^< (\varepsilon) 
- \mbox{\boldmath $\Sigma$}^> (\varepsilon)$,
where $\mbox{\boldmath $\Sigma$}^{<(>)} (\varepsilon)$ 
is the lesser (greater) self-energy for the QD.
$f_N=f(\varepsilon-eV)$,
where $f(\varepsilon)$ is the Fermi distribution function.
Similarly to  $\mbox{\boldmath $\Sigma$}^{r(a)} (\varepsilon)$
in eq. \eqref{eq:Sigma_A},
 the self-energy $\mbox{\boldmath $\Omega$} (\varepsilon)$
can be classified into two parts, 
\begin{eqnarray}
\mbox{\boldmath $\Omega$} (\varepsilon)
=
\mbox{\boldmath $\Omega$}_0 (\varepsilon) 
+ \mbox{\boldmath $\Omega$}_U (\varepsilon),
\label{eq:self_C}
\end{eqnarray}
where $\mbox{\boldmath $\Omega$}_0 (\varepsilon)$ 
($\mbox{\boldmath $\Omega$}_U (\varepsilon)$)
is due to the mixing between the QD and the leads (Coulomb interaction).
Using eq. \eqref{eq:current_C}, we consider the linear conductance
$G_{V=0}=dI/dV|_{V=0}$ at zero temperature.
It should be noted that 
$\textrm{Im}\mbox{\boldmath $\Sigma$}^{r}_U(0)|_{V=0}=0$
at zero temperature,
so that the term including 
$\mbox{\boldmath $\Omega$}_U (\varepsilon)$ vanishes.
Namely, assuming that 
$\textrm{Im}\mbox{\boldmath $\Sigma$}^{r}_U(0)|_{V=0}=0$
and
$\mbox{\boldmath $\Omega$}_U (0)|_{V=0}=0$ at zero temperature,
correlation effects on $G_{V=0}$ at zero temperature enter through 
$\textrm{Re}\mbox{\boldmath $\Sigma$}^{r}_U(0)|_{V=0}$.
On the other hand, the exact formula for
 $\mbox{\boldmath $\Omega$}_0 (\varepsilon)$ is given by
\begin{eqnarray}
\mbox{\boldmath $\Omega$}_0 (\varepsilon)
\!\!\!\!\!&=&\!\!\!\!\! 
-2i
 \left(
   \begin{array}{cc}
    (1-2f_N)\Gamma_N+(1-2f_S)\Gamma_S \tilde{\rho}_S (\varepsilon) \\
    -(1-2f_S)\Gamma_S \tilde{\rho}_S (\varepsilon)
    \frac{\Delta}{\varepsilon} 
   \end{array}
 \right.
\nonumber\\
&& \qquad\quad
 \left.
   \begin{array}{cc}
    -(1-2f_S)\Gamma_S \tilde{\rho}_S(\varepsilon)
     \frac{\Delta}{\varepsilon} \\
    (1-2\bar{f}_N)\Gamma_N+(1-2f_S)\Gamma_S \tilde{\rho}_S (\varepsilon)
   \end{array}
 \right)_,
\nonumber\\
\label{eq:omega0_C}
\end{eqnarray}
where $f_S=f(\varepsilon)$,
 $\bar{f}_N=f(\varepsilon+eV)$ and
  $\tilde{\rho}_S(\varepsilon)=
   \frac{|\varepsilon|\theta(|\varepsilon|-\Delta)}
        {\sqrt{\varepsilon^2-\Delta^2}}_.$
Using the retarded (advanced) Green function in eq. \eqref{eq:Dyson_A}
together with eq. \eqref{eq:omega0_C}, we obtain
the linear conductance at the zero temperature as,
\begin{eqnarray}
G_{V=0}
=
\frac{4e^2}{h} \;
4\Gamma_{N}^2 \;
|\bar{F}_{d\uparrow,d\downarrow}^r (0)|^2_.
\label{eq:conV0_C}
\end{eqnarray}
As shown in eq. \eqref{eq:Green_A},
$\bar{F}_{d\uparrow,d\downarrow}^r (\varepsilon)$
is the (1,2) element of $\textbf{G}_{d,d}^{r} (\varepsilon)$.
Using $G_{\gamma_{-1}\uparrow,\gamma_{-1}\uparrow}^r (\varepsilon )$
in eq. \eqref{eq:Green_ene0} and eqs. 
\eqref{eq:Bogo_BU}, \eqref{eq:Gd_B} and \eqref{eq:Ggamma_B}, we can rewrite
$\bar{F}_{d\uparrow,d\downarrow}^r (\varepsilon)$ around the Fermi energy as
\begin{eqnarray}
\bar{F}_{d\uparrow,d\downarrow}^r (\varepsilon)
\!\!\!\!\!&=&\!\!\!\!\! 
u_dv_d
\left(
G_{\gamma_{-1}\uparrow,\gamma_{-1}\uparrow}^r (\varepsilon )
-
\bar{G}_{\gamma_{-1}\downarrow,\gamma_{-1}\downarrow}^r (\varepsilon )
\right)
\nonumber\\
\!\!\!\!\!&\simeq&\!\!\!\!\!
u_dv_d
\left(
  \frac{z}{\varepsilon-\widetilde{E}_d+i\widetilde{\Gamma}_N}
  -
  \frac{z}{\varepsilon+\widetilde{E}_d+i\widetilde{\Gamma}_N}
\right)_.
\nonumber\\
\label{eq:GreenF_d_C}
\end{eqnarray}
Substituting eq. \eqref{eq:GreenF_d_C} into eq. \eqref{eq:conV0_C},
we have
\begin{eqnarray}
G_{V=0}
=
\frac{4e^2}{h} \;
4\frac{\Delta_d^2}{E_d^2} \;
  \frac{\widetilde{\Gamma}_N^2 \widetilde{E}_d^2}
       {\left( \widetilde{\Gamma}_N^2+\widetilde{E}_d^2 \right)^2}_.
\label{eq:conV0no2_C}
\end{eqnarray}
Note that the factor $\Delta_d^2 / E_d^2$ coming from
$u_d$ and $v_d$ in eq. \eqref{eq:Bogo_factor_B}
is not renormalized by the Coulomb interaction.


\end{document}